\documentclass[10pt,journal,compsoc,twoside]{IEEEtran}

\ifCLASSOPTIONcompsoc
\usepackage[nocompress]{cite}
\else
\usepackage{cite}
\fi

\IEEEoverridecommandlockouts
\usepackage{amsmath,amssymb,amsfonts}
\usepackage{algorithmic}
\usepackage{graphicx}
\usepackage{textcomp}
\usepackage{xcolor}
\usepackage{tabularx,booktabs}
\usepackage{multicol}
\usepackage{url}
\usepackage[skins,breakable]{tcolorbox}
\tcbset{enhanced, sharp corners, colback=gray!15, boxrule=0pt, top=4pt, bottom=4pt, left=4pt, right=4pt,width=\linewidth}

\usepackage{tikz}
\newcommand*\circledC[1]{%
	\tikz[baseline=(C.base)]\node[draw,circle,inner sep=1pt](C) {\scriptsize\textsf{#1}};
}
\newcommand*\circledCC[1]{%
	\tikz[baseline=(C.base)]\node[draw,circle,inner sep=0.2pt](C) {\scriptsize\textsf{#1}};
}

\def\BibTeX{{\rm B\kern-.05em{\sc i\kern-.025em b}\kern-.08em
    T\kern-.1667em\lower.7ex\hbox{E}\kern-.125emX}}

\colorlet{changecolor}{black}

\begin{document}


\title{
Developing Self-Adaptive Microservice Systems: Challenges and Directions
}

\author{Nabor~ C.~Mendon\c{c}a,~
	Pooyan~Jamshidi,~
	David~Garlan,~
	and~Claus~Pahl
	\IEEEcompsocitemizethanks{\IEEEcompsocthanksitem N. C. Mendon\c{c}a is with the 
		Post Graduate Program in Applied Informatics, University of Fortaleza, Fortaleza, CE, Brazil.\protect\\
		E-mail: nabor@unifor.br
		\IEEEcompsocthanksitem P. Jamshidi is with the 
		Computer Science and Engineering Department, University of South Carolina, 	Columbia, SC, USA.\protect\\
		E-mail: pjamshid@cse.sc.edu
		\IEEEcompsocthanksitem D. Garlan is with the 
		School of Computer Science, Carnegie Mellon University,
		Pittsburgh, PA, USA.\protect\\
		E-mail: garlan@cs.cmu.edu
		\IEEEcompsocthanksitem C. Pahl is with the 
		Faculty of Computer Science, Free University of Bozen-Bolzano,
		Bozen-Bolzano, Italy.\protect\\
		E-mail: cpahl@unibz.it		
		}
    \thanks{}}

\markboth{}%
{Mendon\c{c}a \MakeLowercase{\textit{et al.}}: Developing Self-Adaptive Microservice Systems}
%

\IEEEtitleabstractindextext{%
	\begin{abstract}
    A self-adaptive system can dynamically monitor and adapt its behavior to preserve or enhance its quality attributes under uncertain operating conditions. This article identifies key challenges for the development of microservice applications as self-adaptive systems, using a cloud-based intelligent video surveillance application as a motivating example. It also suggests potential new directions for addressing most of the identified challenges by leveraging existing microservice practices and technologies.
	\end{abstract}
	
}
	
\maketitle

\section{Introduction}

A self-adaptive system can monitor its behavior and change its configuration or architecture at run time to preserve or enhance its quality attributes (e.g., performance, reliability, and security) under uncertain operating conditions (e.g., varying workloads, errors, and security threats)~\cite{kephart2003vision}. 
 Despite significant progress in developing self-adaptive systems in recent years, only a handful of techniques, such as automated server management, cloud elasticity, and automated data center management, have thus far found their way to industrial applications~\cite{weyns2017software}.
 For example, Kubernetes,\footnote{\url{https://kubernetes.io/}} a modern container orchestration platform that is increasingly being used to deploy and \textcolor{changecolor}{manage microservice} applications in the cloud~\cite{microservicesjourney2018} only provides autoscaling, i.e., the ability to automatically change the number of instances of a service and self-healing, i.e., the ability to automatically restart failed service instances, as part of its native self-adaptation capabilities.

The popularity of microservices in industry has naturally sparked the interest of the research community. However, most existing research on microservices \textcolor{changecolor}{focuses} on general architectural principles and migration guidelines (e.g.,~\cite{balalaie2016microservices}). Only a few works have addressed the specific challenges of developing microservice applications as self-adaptive systems; nevertheless, even those works tend to be rather narrow in scope, offering only limited forms of adaptation (e.g., self-healing~\cite{rajagopalan2015app} and run-time placement adaptation~\cite{sampaio2019adaptation}).

We believe the uptake of microservices and their enabling practices and technologies by industry brings a unique opportunity to narrow the gap between the state-of-the-art and the state-of-the-practice in self-adaptive systems, and that both self-adaptive systems and microservice communities have much to gain from each other.

On the one hand, recent progress in self-adaptive systems offers a control-oriented perspective that leverages a large body of theoretical and practical results to enhance microservice quality attributes using techniques such as planners (e.g., to select the best possible adaptation strategy for each microservice), machine learning (e.g., to learn new adaptation strategies from past adaptation results), reasoning under uncertainty (e.g., to cope with noisy monitoring data), and multi-objective optimization (e.g., to cater to multiple, possibly conflicting microservice requirements)~\cite{filieri2017control}. On the other hand, software characteristics such as independent and frequent deployment, the need for a high degree of automation, and complex run-time architectures~\cite{microservicesjourney2018} make microservices a fertile ground to foster further research and development on self-adaptive systems.

In this paper, we take a closer look into the interplay between self-adaptive systems and microservices in order to identify key challenges for the development of microservice applications as self-adaptive systems. Our contributions are as follows: (1) we describe a cloud-based intelligent video surveillance application as an example of a self-adaptive microservice system; (2) we identify and illustrate, in the context of this example application, several challenges for microservice development, delivery, and operations from multiple self-adaptation perspectives; and (3) we discuss potential new directions for addressing most of the key challenges identified for developing self-adaptive microservice systems by leveraging existing microservice practices and technologies.

\begin{figure*}[t]
	\centering
    \includegraphics[width=.95\textwidth]{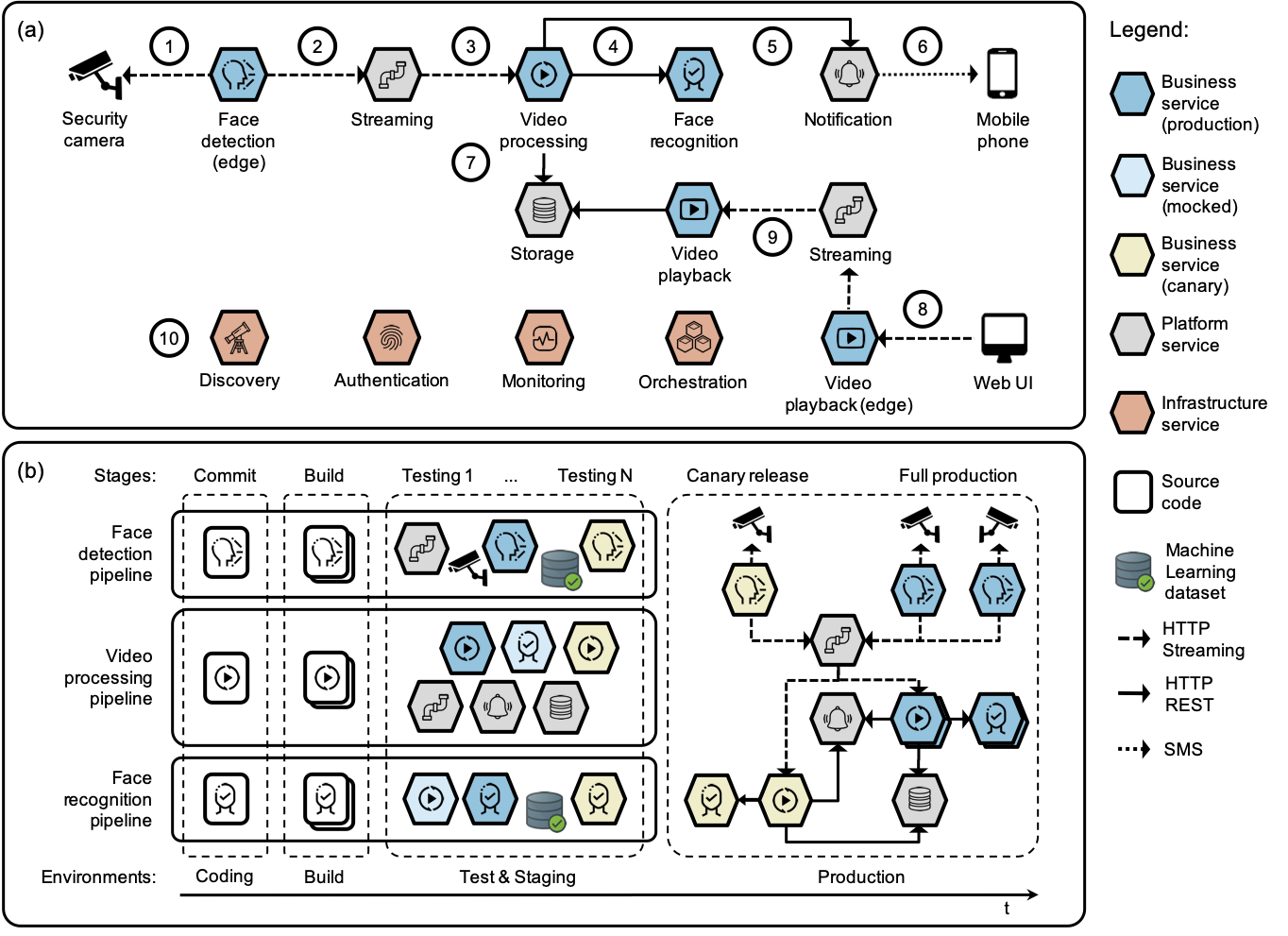}
	\caption{(a) Microservice architecture for the edge-cloud intelligent video surveillance application; and (b) three continuous delivery pipelines for the application's face detection, video processing, and face recognition microservices, respectively (adapted from~\cite{heinrich2017performance}).}
	\label{fig:ms-arch-cd-pipelines}
\end{figure*}

\section{Microservices}

Microservices constitute an emerging architectural style that builds on the well-established concept of modularization but emphasizes technical boundaries (i.e., different address and execution spaces) between software components~\cite{microservicesjourney2018}. Each module---each microservice---is developed around a single business capability that offers access to its internal logic and data through a well-defined network interface. Due to their relative simplicity and small size, microservices can be released often and adapted  to different production environments--- from cloud data centers to resource-constrained devices (e.g., IoT)---without much development effort. This has a significant impact on improving software agility, because each microservice becomes an independent unit of development, deployment, versioning, scaling and management~\cite{microservicesjourney2018}.

Microservices are considered to be an enabler for emerging DevOps practices, such as continuous integration (CI) and continuous delivery (CD), which aim to significantly decrease the time between changing a system and transferring that change to the production environment~\cite{devops:book}. To achieve this level of agility, microservices are typically packaged in lightweight containers (e.g., Docker\footnote{\url{https://www.docker.com/}}) and deployed and managed using automated container orchestration tools (e.g., Kubernetes). In particular, the use of fully automated CD tools (e.g., Spinnaker\footnote{\url{https://www.spinnaker.io/}}) enables the creation of independent CD pipelines for each microservice. 
Having multiple CD pipelines running in parallel allows microservice changes to be rapidly delivered into production since only the affected microservice needs to be built and tested~\cite{heinrich2017performance}.

\section{A Self-Adaptive Microservice System}\label{sec:exemplar}

We consider an edge-cloud intelligent video surveillance application as an example of a self-adaptive microservice system. The application's goal is to alert users about the presence and possibly the identification of humans in a certain location via real-time analysis of video frames captured by one or more security cameras. Its architecture, inspired by one of Amazon's machine learning (ML) sample solutions,\footnote{\url{https://aws.amazon.com/blogs/machine-learning/create-a-serverless-solution-for-video-frame-analysis-and-alerting/}} is composed of several \emph{business}, \emph{platform}, and \emph{infrastructure} services, as shown in Figure~\ref{fig:ms-arch-cd-pipelines}(a).

Each security camera is deployed along with a ML-based \emph{face detection} edge service.$^{\circledC{1}}$ This service is a less accurate yet more energy efficient version of a ML-based \emph{face recognition} service deployed in the cloud. The role of the face detection edge service is to select relevant video frames (i.e., frames with detected human faces) to be asynchronously transmitted to a \emph{video processing} service in the cloud using a cloud-provided \emph{streaming} service.$^{\circledC{2}}$ 
The streaming service offers a more scalable and reliable solution over the basic HTTP Streaming protocol. The video processing service in the cloud analyzes the received video frames\;$^{\circledC{3}}$ and passes them as invocation parameters to the face recognition service$^{\circledC{4}}$ in an attempt to recognize any person that might be visible in them. If the face recognition service reports a match, the video processing service sends out an SMS message informing the application users' \emph{mobile phones} about the match using a \emph{notification} service.$^{\circledC{5}\circledC{6}}$ The video processing service then saves the analyzed video frames along with any relevant information regarding the video recording (e.g., date, camera ID) and contents (e.g., names of the recognized persons, match accuracy) in a \emph{storage} service.$^{\circledC{7}}$ Application users can use a \emph{Web-based User Interface} (Web UI) to search for and play back any analyzed video segment stored in the cloud. This is done by invoking a \emph{video playback} edge service deployed somewhere closer to the user's physical location.$^{\circledCC{8}}$ The video playback edge service, in turn, uses the streaming service to communicate asynchronously with a \emph{video playback} service in the cloud.$^{\circledCC{9}}$ 

The video surveillance application also relies on several \emph{infrastructure services}, such as a \emph{discovery} service, an \emph{authentication} service, a \emph{monitoring} service, and a \emph{container orchestration} service.$^{\circledCC{10}}$
To avoid cluttering, the depiction of the video surveillance application architecture in Figure~\ref{fig:ms-arch-cd-pipelines}(a) omits interactions between business services and infrastructure services.


Following typical microservice practices, each service of the video surveillance application is delivered in production using its own independent CD pipeline. Figure~\ref{fig:ms-arch-cd-pipelines}(b) depicts three possible CD pipelines for the face detection, video processing, and face recognition services. Note how each pipeline goes through different environments, from coding to production, which represent the multiple development stages (e.g., build, testing, canary release, and finally, full production) of each service~\cite{devops:book}. 

We envision multiple self-adaptation scenarios for this example application, including the following:

\begin{itemize}
	\item The video processing service may need to dynamically change its number of deployed instances in response to load variations;
	
	\item The \textcolor{changecolor}{face recognition} service may need to dynamically change its container image (e.g., to switch to a less accurate version) under extreme load conditions;
	
	\item The video playback service may need to dynamically change its quality attributes (e.g., its frame rate in order to cope with latency fluctuations); and
	
	\item The CD pipeline for the video playback edge service may need to dynamically adjust that service's testing parameters (e.g., to expedite the testing of its self-healing capabilities during staging).
\end{itemize}


Enacting all of the foregoing different scenarios may bring up a number of challenges for microsevice application developers, which we discuss next.

\section{Challenges}\label{sec:challenges}

We identify the main challenges facing the development of self-adaptive microservice systems from four perspectives: \emph{design space}, \emph{control loop deployment}, \emph{continuous delivery}, and \emph{testing}. 

\subsection{Design Space} \label{sec:design-space}

Designing self-adaptive systems involves making design decisions about the environment while it is being observed and about the system itself, and then selecting adaptation mechanisms that are thereafter enacted~\cite{weyns2017software}. In the context of a microservice application, the design space for making self-adaptation decisions is even more complex, due to the large number of run-time components and their independent and highly dynamic nature~\cite{microservicesjourney2018}. Thus, a first challenge is as follows: 

	\begin{tcolorbox}[]
			\textbf{[C1]}~How to determine monitoring and adaptation mechanisms to face the diversity of microservices' quality attributes.
	\end{tcolorbox}

The  case of the example video surveillance application illustrates this challenge as the quality requirements and adaptation needs of the ML face recognition service, which handles only individual video frames and is invoked synchronously from within the cloud, might be quite different from those of the video playback edge service, which handles the entire video segments and is invoked asynchronously from outside the cloud. In particular, the former would not have to monitor and adapt its communication parameters (e.g., the request rate) at run time; in contrast, this  would be a critical requirement for the latter.

An additional challenge is the following: 

	\begin{tcolorbox}[]
			\textbf{[C2]}~How to identify and resolve potential conflicts between the quality requirements of individual microservices and those of the overall application when defining their self-adaptive behaviors.
	\end{tcolorbox}

For instance, an important quality attribute of a face recognition service, which typically works by comparing extracted facial features from a given image with facial images within a database, is its recognition accuracy. However, high accuracy in this kind of ML service inevitably implies high processing and storage costs, which might conflict with the application's overall cost constraints. 

The following is yet another challenge in this context:

	\begin{tcolorbox}[]
			\textbf{[C3]}~How to reconcile the adaptation needs of individual microservices and the overall application with the self-adaptation capabilities offered by the underlying infrastructure management platform.
	\end{tcolorbox}

By way of illustration, while Kubernetes' autoscaling capabilities could be useful to improve the response time of all cloud-based services deployed as part of the video surveillance application, current Software Reliability Engineering (SRE) practices indicate that autoscaling alone might not be enough to make strong guarantees about a service's performance, especially under high load.\footnote{\url{https://landing.google.com/sre/}} In addition, Kubernetes does not yet support monitoring and management of service communication characteristics at the application layer (i.e., Layer~7), such as by establishing per service rate limits and bandwidth quotas, which could be useful to help manage the performance and communication overhead of the video playback service. 


Finally, the more recent Function-as-a-Service (FaaS) and serverless computing cloud models,\footnote{\textcolor{changecolor}{Mike Roberts (May 22, 2018). \emph{Serverless Architecture}. MartinFowler.com. Retrieved November 13, 2019, from \url{https://martinfowler.com/articles/serverless.html}}} in which fined-grained services are deployed as serverless functions that are transparently managed and scaled by the cloud platform and charged on a per usage basis, poses an additional challenge:

	\begin{tcolorbox}[]
			\textbf{[C4]}~How to develop and manage self-adaptive microservice systems in hybrid deployment environments that are composed of both serverful and serverless services.
	\end{tcolorbox}

A case in point is encountered because some of the video surveillance services (e.g., video processing) could be re-implemented and re-deployed as serverless services using an FaaS cloud platform, such as AWS Lambda.\footnote{\url{https://aws.amazon.com/lambda/}} In that case, application developers would not only have to reconcile the adaptation requirements of both serverful and serverless services with the self-adaptation capabilities of their respective management platforms but also to rethink their entire \mbox{DevOps} and business strategies in order to account for the significant technical and economical differences between those two deployment models~\cite{adzic2017serverless}.

\subsection{Control Loop Deployment} \label{sec:control}

Control loops are crucial elements necessary to realize the run-time adaptation of software systems. A typical self-adaptation control loop consists of four main activities, namely Monitor, Analyze, Plan, and Execute, which share a common Knowledge base, that are usually referred to as the MAPE-K reference model~\cite{kephart2003vision}. Control loops can be designed and deployed according to different control strategies---from a single centralized control component managing the whole system to multiple control components managing different parts of the system and organized in a hierarchical or fully decentralized manner~\cite{weyns2017software}. In the context of a microservice application, where each microservice is independently developed, deployed, and managed at run time, selecting the appropriate control strategies poses additional challenges. 

The following constitutes an initial challenge in this context:

	\begin{tcolorbox}[]

		\textbf{[C5]}~How to determine the level of \emph{distribution}, \emph{visibility}, and \emph{granularity} necessary for deploying a microservice application's control components.

	\end{tcolorbox}

We see these three deployment dimensions as forming a control loop's \emph{deployment space} for a microservice application. The distribution dimension concerns the physical allocation of control loop components to infrastructure resources. The visibility dimension, in turn, concerns whether the control loop components should be deployed at the application level (i.e., fully visible to developers) or at the infrastructure  level (i.e., only partially visible to developers). Finally, the granularity dimension concerns whether control loop components should be deployed as a single monolithic service or decomposed into a collection of independently developed and managed microservices.  

Developing and deploying self-adaptive microservice systems with those three dimensions in mind could help to establish fundamental tradeoffs with respect to multiple quality attributes. An example is deploying control loops for each of the video surveillance application services in a fully decentralized fashion, which would in turn increase their overall reliability and scalability. Nevertheless, this strategy would also make it harder to enforce application-wide adaptation constraints, as this would require each control loop to coordinate its actions with the other control loops, thereby reducing their autonomy. Similarly, deploying control loops at the infrastructure level would help to promote a better separation between business and management services at run time, thus facilitating their reuse. Despite this benefit, this strategy would also make control loops much harder to customize for specific adaptation needs (e.g., managing the expected accuracy of ML services), as most control loops deployed at the infrastructure level support only a restricted set of adaptation models and mechanisms~\cite{mendonca2018generality}. Finally, deploying control loop components as monolithic services would greatly simplify their packaging and management at run time. However, this strategy would also force all control components to share the same version and release rate, thus severely compromising their continuous improvement to satisfy evolving adaptation needs.

Another challenge related to control loop deployment is as follows:

	\begin{tcolorbox}[]
		\textbf{[C6]}~How to reconcile the selected control loop deployment strategies with those supported by the underlying infrastructure management platform.
	\end{tcolorbox}

As an example, in Kubernetes, self-healing and autoscaling controllers are deployed in a mostly centralized manner, as part of its master node. Therefore, if each microservice is to be managed by a fully independent control loop, there should be multiple instances of such types of controllers deployed, one for each microservice. In addition, those individual controllers might still have to coordinate their actions at the application level in order to resolve potential requirement conflicts, as discussed in Section~\ref{sec:design-space}, which coordination Kubernetes currently does not support. 

\subsection{Continuous Delivery}
\label{sec:cd-challenges}

We identify three main adaptation scenarios concerning the use of CD practices and tools in the context of a self-adaptive microservice system. The first scenario is the run-time adaptation of the CD pipelines themselves. In that regard, the following is a key challenge: 

\begin{tcolorbox}[]
	\textbf{[C7]}~How to best determine the appropriate monitoring and adaptation mechanisms that will dynamically adjust the transition events and conditions across the stages of each microservice CD pipeline.
\end{tcolorbox}

An exemplar of this challenge is as follows: To create a CD pipeline developers typically need to define the events and conditions under which each stage of the pipeline can be executed. This is usually done by manually expressing the trigger events and test requirements of each stage so that the target application can be automatically progressed from one stage to the next. In this scenario, a self-adaptive pipeline would have to be able to monitor the application behavior at each stage so as to dynamically adjust its trigger events \mbox{and/or} test requirements to cope with uncertainties during its execution (e.g., the application repeatedly failing to meet the requirements of one particular stage due to the necessary data or infrastructure resources being unavailable).

Yet another challenge related to this scenario is as follows:

\begin{tcolorbox}[]
	\textbf{[C8]}~How to reconcile the adaptation needs of each microservice CD pipeline with the self-adaptation capabilities offered by the underlying infrastructure and CD management platforms.
\end{tcolorbox}

Kubernetes offers a rollout operation that can be used to illustrate this challenge. A rollout operation automatically deploys a new service version without having to take the current version down. This operation could be useful to implement a \emph{canary release} pipeline for each of the video surveillance services. A canary release pipeline is responsible for automatically deploying a new service version to production in parallel with the most recent stable version of that service so as to gradually replace instances of the stable service version with instances of the new (canary) version based on a given set of testing criteria~\cite{devops:book}. However, this solution would still require service developers to manually invoke Kubernete's rollout operation every time there's a new service version to be released. Alternatively, developers could benefit from a full-fledged CD management tool (e.g., Spinnaker) to fully automate the execution of that canary release pipeline. But even in that case, developers would still need to provide the CD management tool with the exact conditions and their trigger events required to deploy the canary release into production. 

The second scenario is the adaptation of the microservices' own adaptation requirements. In this scenario, a key challenge faced is the following demand:

\begin{tcolorbox}[]
	\textbf{[C9]}~How to dynamically adjust the adaptation requirements of each microservice according to the needs of each CD stage.
\end{tcolorbox}

As an example, to streamline the testing of the autoscaling capabilities of the video processing service, the CD platform could automatically adjust that service's performance requirements during staging so as to trigger its autoscaling features earlier and more often than during normal production. 

Finally, the third scenario concerns treating the self-adaptation mechanisms of a self-adaptive microservice system as first-class DevOps entities. In that respect, a further challenge is addressing the following issue:

\begin{tcolorbox}[]
	\textbf{[C10]}~How to establish an effective strategy for continuously delivering self-adaptation mechanisms in production in the context of a self-adaptive microservice system.
\end{tcolorbox}

By way of example, each self-adaptation mechanism used by the video surveillance application (e.g., autoscaling and self-healing) could be independently tested and deployed in its own CD pipeline. Such a strategy would allow a more systematic reuse of self-adaptation models and mechanisms across business services, albeit at the expense of having a more complex CD process.

\subsection{Testing}

Conducting extensive validation tests with microservices before each deployment is not feasible due to the high frequency of their releases. Instead, microservice quality assurance is often compensated or even replaced by fine-grained monitoring techniques in production environments exposed to real workloads. In this way, failures can be monitored and quickly corrected by pushing new releases into production~\cite{heinrich2017performance}.

The following is a key challenge in this context:

	\begin{tcolorbox}[]
		\textbf{[C11]}~How to determine testing models and mechanisms to assess the fundamental quality attributes of each microservice and of the overall application from a self-adaptation perspective.
	\end{tcolorbox}

An example of this challenge in the context of the video surveillance application is that developers may need to select different testing mechanisms to assess the self-adaptation features of different microservices (e.g., image benchmarking and load generation for testing the autoscaling features of the face recognition service and fault injection for testing the self-healing features of the video playback service). In addition, the need for those testing mechanisms may vary across CD stages and pipelines (e.g., autoscaling tests of the face recognition service may be run only during staging while more critical self-healing tests of the video playback service may be run all the way to production).

An additional challenge related to testing is the following:

	\begin{tcolorbox}[]
		\textbf{[C12]}~How to integrate the systematic testing of microservices, including their self-adaptive behaviors, within the context of existing CD practices.
	\end{tcolorbox}

The use of fault injection mechanisms to test a self-adaptive system's resiliency in production illustrates this challenge. For example, the results of fault injection tests created to assess the resiliency of the self-healing features of the video playback service could be used by the underlying CD platform as one of the criteria for deciding whether the latest version of that service is ready to be promoted from canary release to normal production.

\section{New Directions}\label{sec:directions}
We now suggest promising new directions to address most of the challenges identified in the previous section. 

\subsection{New Adaptation Mechanisms}

Regarding challenges C1 and C3, a practical way of implementing novel self-adaptation solutions tailored for the context of microservice applications would be by extending the management API provided by current container orchestration tools, like Kubernetes. To illustrate, one could easily build on the rollout and rollback features provided by Kubernetes to develop a self-adaptive service fallback mechanism that could be customized to meet different adaptation requirements. 

The basic idea is to create multiple fallback versions for each microservice (e.g., a low fidelity version or a high accuracy version) and pack them as separate Docker images. The self-adaptive fallback mechanism could then be configured to use Kubernetes to automatically update the current image of a given microservice to one of its fallback images under certain system or environment conditions (e.g., update the face recognition service to its high accuracy image whenever the system is underutilized and rollback that service to its original image whenever the load reaches a certain threshold). 

Similarly, one could build on the variety of communication-related monitoring and management features provided by the use of so-called \emph{service mesh} tools~\cite{microservicesjourney2018} to create novel connector-oriented self-adaptation mechanisms. This is the case, for example, of using the security management features of a service mesh tool like Istio\footnote{\url{https://istio.io/}} to provide a self-protection mechanism that will automatically strengthen the encryption parameters of the communication protocols being used by the video surveillance application services under a security attack, without the need to change or restart any service.

A service mesh could also be useful to dynamically adjust the \emph{retry and circuit-breaking} parameters~\cite{microservicesjourney2018} used by all clients of an overloaded service so as to reduce the service's incoming traffic and thus prevent cascading failures from propagating throughout the system. This feature could be particularly useful for mission-critical self-adaptive systems.

Finally, specifically regarding challenge C2, application developers may rely on recent microservice orchestration languages \textcolor{changecolor}{(e.g., Jolie\footnote{\textcolor{changecolor}{\url{https://www.jolie-lang.org/}}})} as the means for defining and enacting application-aware adaptations. While this may provide a possible solution, its adoption may be seen as counterintuitive with respect to some well-established microservice principles, such as the autonomy of each microservice team to choose the implementation technologies that best fit their needs and expertise~\cite{microservicesjourney2018}.

\subsection{New Control Loop Deployment Structures}

From a control loop deployment perspective, as discussed in the context of challenges C5 and C6, a microservice system's control components should ideally be deployed in a fully decentralized fashion, with each microservice being managed by its own local controller. The downside of this solution is that it makes it harder for the local controllers to monitor and manage application-wide quality attributes. Although having a centralized controller dedicated to managing application-level quality concerns would be a more straightforward solution for this issue, its use would create a single point of failure and ultimately could compromise the application's overall availability. In practice, microservice developers may choose from a variety of intermediate solutions between those two extremes, e.g., by logically grouping services according to their business and/or quality  affinity and then having those services being collectively managed by independent yet application-aware group controllers organized in a hierarchical or fully decentralized structure.

Another important issue is the decision about whether microservice developers should have any responsibility in developing and managing control components, as suggested above. 
Having control components explicit in the design and development of a self-adaptive microservice system may contribute to further increase the system's overall complexity. In addition, this decision could compromise the delivery autonomy of individual services, since service developers would have to negotiate before every new release in case their code bases share the same control components. We advocate the adoption of a middle ground solution: having all control components developed, tested, and released as part of the underlying infrastructure management platform yet still exposed to service developers during the later phases of their services' CD pipeline~\cite{mendonca2018generality}.     

\subsection{New Continuous Delivery  Strategies}

The question of how to integrate self-adaptation capabilities with DevOps, discussed in the context of challenges C7--C9, deserves special attention from self-adaptive microservice system developers. Here we discuss two possible CD strategies and their trade-offs.

One first strategy is to create a new dedicated CD pipeline to deliver into production all self-adaptation components used by the application. Having a dedicated self-adaptation pipeline would have the advantage of running it in parallel with the other business pipelines, and thus preserve their autonomy. On the other hand, this solution would prevent business developers from directly improving the self-adaptation components used by their services, as this would require negotiating with the developers responsible for the self-adaptation pipeline. Moreover, this solution would push critical integration tests to the end of each business pipeline, as business developers would not have immediate access to the latest version of their required self-adaptation components.

A second strategy is to have separate business and self-adaptation pipelines during the commit and build stages, but have the self-adaptation pipeline integrated with the other business pipelines as soon as they enter the testing stage. \textcolor{changecolor}{In this case, self-adaptation components would still be designed and developed by a separate team, but their testing and deployment would be the responsibility of business developers, as part of their business service pipelines.} This would \textcolor{changecolor}{have the benefit of avoiding} delaying integration tests in the business pipelines. However, as a side effect, it could unnecessarily bloat all business pipelines, with business developers now being responsible for testing both business and self-adaptation components.

Those two strategies are but a small sample of the spectrum of CD alternatives developers of self-adaptive microservice systems can explore. In that direction, extending emerging Model-Driven Engineering (MDE) tools for microservices (e.g., AjiL\footnote{\url{https://github.com/SeelabFhdo/AjiL}}) with CD support could be a key enabler to automate the integration of self-adaptation capabilities and DevOps.

\subsection{New Testing Approaches}

A number of recent testing approaches have been proposed specifically for microservices both by industry (e.g., integration testing in production\footnote{\url{https://labs.spotify.com/2018/01/11/testing-of-microservices/}}) and by academia (e.g., trace-based error prediction and fault localization~\cite{zhou2019}). A natural new direction here, related to challenges C11 and C12, is to systematically integrate those emerging testing approaches within the context of a fully automated self-adaptive CD pipeline. In this way, the results of both online and offline tests could be incorporated as part of the set of dynamic events and conditions used by \textcolor{changecolor}{the} underlying CD platform to automatically test a given microservice release across multiple pipeline stages.

Yet another promising direction in this context is applying \emph{chaos engineering}~\cite{basiri2016chaos} principles to assess the \emph{resiliency} of  a self-adaptive microservice system~\cite{camara2017robustness}.
In that regard, one could rely on such a tool as kube-monkey,\footnote{\url{https://github.com/asobti/kube-monkey}} which can be used to randomly terminate service instances in a Kubernetes cluster, to develop a kind of \emph{adversarial control loop}, whose main goal is to disturb the self-adaptive behavior of a target microservice system. This adversarial control loop could then leverage existing self-adaptation models and techniques to monitor and respond to (i.e., counteract) the microservice system's adaptation actions, and thereby offer a powerful mechanism to test the self-adaptive system's resiliency in production. 

\subsection{\textcolor{changecolor}{New Migration Strategies to Microservices}}

\textcolor{changecolor}{
In recent years the migration from legacy monolithic applications to a microservice-based architecture has gained considerable momentum, in both industry~\cite{monolith-microservices:book} and academia~\cite{balalaie2016microservices}. However, questions such as when and how to switch to the newly migrated microservice system are still challenging. One popular migration strategy is applying the Strangler Fig Application pattern,\footnote{\textcolor{changecolor}{\url{https://martinfowler.com/bliki/StranglerFigApplication.html}}} in which the functionalities implemented by the monolithic application are gradually replaced by new microservices. In that regard, an interesting new direction is to use a self-adaptive service mesh to dynamically reroute traffic from the monolith to the new services as they are delivered into production. This would facilitate the task of introducing and testing new services during the migration process, and also enable a gradual decommission of the legacy infrastructure.   
}

\section{Conclusion}\label{sec:conclusion}

We have looked into a number of practical challenges facing the development and delivery of self-adaptive microservice systems. We have also suggested potential ways to address most of those challenges, with a focus on current and emerging microservice practices and technologies. We hope our ideas contribute to promote a better understanding of the interplay between self-adaptive systems and microservices, thus providing a timely incentive for researchers, practitioners, and tool developers to tackle some of the issues raised here as well as others that might arise. 

\ifCLASSOPTIONcompsoc


\ifCLASSOPTIONcaptionsoff
\newpage
\fi



\bibliographystyle{IEEEtran}
\bibliography{IEEEabrv,references}


%
\vspace{-25pt}

\begin{IEEEbiography}[{
        \includegraphics[scale=.4,clip,keepaspectratio]{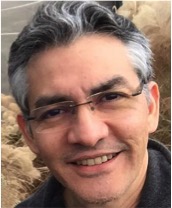}
		}]{Nabor C. Mendon\c{c}a}
	 is a professor of applied informatics at the University of Fortaleza, Brazil. From 2017 to 2018 he was a visiting scholar in the School of Computer Science at Carnegie Mellon University, US. His research interests include software engineering, distributed systems, self-adaptive systems, and cloud computing. Mendon\c{c}a received a Ph.D. in computing from Imperial College London. 
     Contact him at nabor@unifor.br.
\end{IEEEbiography}


\vspace{-25pt}

\begin{IEEEbiography}[{
        \includegraphics[scale=.4,clip,keepaspectratio]{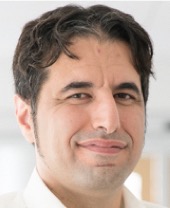}       
	}]{Pooyan Jamshidi}
	is an assistant professor at the University of South Carolina, US. His research interests include software engineering, systems, and machine learning, with a focus on the areas of machine-learning systems. Jamshidi received a Ph.D. from Dublin City University, Ireland. Contact him at pjamshid@cse.sc.edu.
\end{IEEEbiography}



\begin{IEEEbiography}[{
        \includegraphics[scale=.4,clip,keepaspectratio]{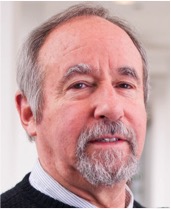}
	}]{David Garlan}
	is a professor and associate dean in the School of Computer Science at Carnegie Mellon University, US. His research interests include autonomous and self-adaptive systems, software architecture, formal methods, explainablity, and cyberphysical systems. Garlan received a Ph.D. in computer science from Carnegie Mellon University. Contact him at \mbox{garlan}@cs.cmu.edu.
\end{IEEEbiography}

\newpage

\begin{IEEEbiography}[{
        \includegraphics[scale=.4,clip,keepaspectratio]{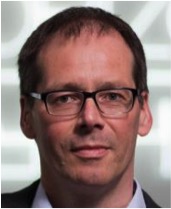}
		}]{Claus Pahl}
	is a professor of computer science at the Free University of Bozen-Bolzano, Italy, where he heads the Software and Systems Engineering Group. His research interests include software engineering in service and cloud computing, specifically migration, architecture specification, dynamic quality, performance engineering, and scalability. Pahl received a Ph.D. in computing from the University of Dortmund. Contact him at cpahl@unibz.it.
\end{IEEEbiography}

%








\end{document}